%% file: main.tex
\documentclass[twocolumn]{aastex631}

\usepackage{amssymb,amsmath}
\usepackage{braket}
\usepackage{longtable}
\usepackage{threeparttable}
\usepackage{threeparttablex}
\usepackage{multirow}
\usepackage{natbib}
\usepackage{hyperref}
\usepackage{array}
\usepackage{afterpage}
\usepackage{graphicx}
\usepackage{bm}
\usepackage{multirow}
\usepackage{lmodern}

\newcommand{\UCI}{Department of Physics and Astronomy, 4129 Frederick Reines Hall, University of California, Irvine, Irvine, CA, 92697, USA}

\newcommand{\PSUAA}{Department of Astronomy \& Astrophysics, 525 Davey Laboratory, The Pennsylvania State University, University Park, PA, 16802, USA}
\newcommand{\PSUCEHW}{Center for Exoplanets and Habitable Worlds, 525 Davey Laboratory, The Pennsylvania State University, University Park, PA, 16802, USA}
\newcommand{\PSETI}{Penn State Extraterrestrial Intelligence Center, 525 Davey Laboratory, The Pennsylvania State University, University Park, PA, 16802, USA}

\newcommand{\LUP}{Department of Physics, Lehigh University, 16 Memorial Drive East, Bethlehem, PA, 18015, USA}
\newcommand{\CFA}{Center for Astrophysics | Harvard and Smithsonian, MS 58, 60 Garden Street, Cambridge, MA 02138, USA}
\newcommand{\TSU}{Center of Excellence in Information Systems, Tennessee State University, Nashville, TN 37209  USA}

\newcommand{\logRHK}{\log{R'_{\mathrm{HK}}}}
\newcommand{\SHK}{$S_{\mathrm{HK}}$}

\newcommand{\CaHK}{\ion{Ca}{2} H \& K}

\definecolor{todo}{rgb}{0.4, 0.0, 0.6}

\newcommand{\comment}[1]{}

\defcitealias{Baum2022}{B22}

\begin{document}

\title{HD 166620: Portrait of a Star Entering a Grand Magnetic Minimum}

\author[0000-0002-4927-9925]{Jacob K. Luhn}
\affiliation{\UCI}

\author[0000-0001-6160-5888]{Jason T.\ Wright}
\affil{\PSUAA}
\affil{\PSUCEHW}
\affil{\PSETI}

\author{Gregory W. Henry}
\affiliation{\TSU}

\author[0000-0001-7032-8480]{Steven H.\ Saar}
\affiliation{\CFA}

\author[0000-0002-9021-9780]{Anna C. Baum}
\affiliation{\LUP}


\email{jluhn@uci.edu}

\begin{abstract}
HD 166620 was recently identified as a Maunder Minimum candidate based on nearly 50 years of \CaHK{} activity data from Mount Wilson and Keck-HIRES \citep{Baum2022}.  These data showed clear cyclic behavior on a 17-year timescale during the Mount Wilson survey that became flat when picked up later with Keck-HIRES planet-search observations. Unfortunately, the transition between these two data sets---and therefore the transition into the candidate Maunder Minimum phase---contained little to no data. Here we present additional Mount Wilson data not present in \citet{Baum2022} along with photometry over a nearly 30-year baseline that definitively trace the transition from cyclic activity to a prolonged phase of flat activity. We present this as conclusive evidence of the star entering a grand magnetic minimum and therefore the first true Maunder Minimum analog. We further show that neither the overall brightness nor the chromospheric activity level (as measured by \SHK) is significantly lower during the grand magnetic minimum than its activity cycle minimum, implying that anomalously low mean or instantaneous activity levels are not a good diagnostic or criterion for identifying additional Maunder Minimum candidates. Intraseasonal variability in \SHK, however, \emph{is}  lower in the star's grand minimum; this may prove a useful symptom of the phenomenon.  

\end{abstract}

\section{Introduction}\label{sec:introduction}

The Maunder Minimum was a period of extraordinarily low sunspot levels from roughly 1645--1715 \citep{Eddy76}. The nature of this apparent pause in the Sun's 11-year sunspot cycle has implications for the nature of the Solar dynamo and for our interpretations of sunspot records in other stars.

Data from the Mount Wilson HK survey \citep{Baliunas95b} and later surveys revealed a population of apparently Sun-like stars with low and constant levels of activity, interpreted by \citet{Baliunas90}, \citet{Saar92}, \citet{Henry96}, and others as ordinarily-cycling stars caught in a Maunder-minimum-like state, or ``grand magnetic minimum.''

\citet{Wright04b}, however, showed that most of these stars are actually slightly evolved, implying that they are not in extraordinary states of low activity amidst longer periods of cycling, but old stars that have stopped cycling entirely as their dynamos die out.  

In the model of \citet{Metcalfe2017}, stars with mean activity levels near a threshold $\log{R^\prime_\mathrm{HK}}$ value of -4.95 \citep{Brandenburg2017} will episodically experience grand minima, a behavior that increases in frequency as the star's mean activity level drops, until it eventually becomes permanent.

\citet{Saar2012} has a good discussion of the difficulties in identifying true Maunder Minimum analog stars and efforts to overcome these difficulties by studying the activity level, activity variability, and evolutionary state of stars. \citet{Donahue1995} identified HD 3651 as a candidate, based on the apparent weakening of its activity cycle, perhaps towards an extended long state.

\citet{Shah18} showed that the weakening of the cycle of HD 3651 had not continued into the 2010's but identified another star, HD 4915, that might be a better candidate because it showed three consecutively weaker activity maxima across 12 years of data. 

In the opposite vein, \citet{Mittag2019} observed an increasing trend in activity-cycle amplitude in HD 140538. \citet{Mittag2019} point to this (and the slightly decreasing trend in the final years of data) as evidence of a longer $\sim$30-year activity cycle on top of the 3.88-year cycle. However, the earliest few observations from the Solar-Stellar Spectrograph (SSS) fall below the expected cycle maximum. Therefore, another plausible interpretation of the SSS data---which otherwise shows good agreement with other contemporaneous data sets---is that of a star caught \emph{exiting} a magnetic grand minimum state.

More recently, \citet{Baum2022}, hereafter \citetalias{Baum2022}, combined Mount Wilson HK project measurements with two sets of Keck-HIRES planet search measurements to show that HD 166620 was once a strongly cycling star but since 2004 has low and constant levels of activity, a stark and dramatic change in behavior that would seem to be unambiguous evidence that it was in a Maunder-Minimum-like state.  

The dramatic change in behavior is perfectly coincident with a gap in observations between Mount Wilson and the upgraded HIRES instrument, inspiring \citetalias{Baum2022} to thoroughly explore and reject the possibility of a mismatch or error in the identification of the star across the two projects.  

Further astrophysical interpretation of the data set is also somewhat complicated by a potential calibration mismatch among the three sets of measurements: most stars in \citetalias{Baum2022} appear to be well calibrated, but a few required adjustments. It is thus unclear if the present-day activity level of the star is truly similar to its ordinary cycle minimum observed by Mount Wilson, or could be at a different level.

Here, we present two additional sets of data that help bridge the gap among the \citetalias{Baum2022} data sets and trace out the entrance into a grand magnetic minimum. The first is some published Mount Wilson data not considered by \citetalias{Baum2022} showing the completion of the star's final cycle before the grand minimum, and the second is new optical photometry during both the end of the final cycle and the present-day grand minimum.

These data both confirm the reality of the transition and suggest the calibration in \citetalias{Baum2022} is very good for this star, meaning that the star's grand minimum activity level is similar to that of its last few cycle minima.

\section{Data}
\subsection[]{\SHK{} activity from Mount Wilson \& Keck-HIRES \citep{Baum2022}}
We primarily use the time series given in \citetalias{Baum2022} that initially identified HD 166620 as a Maunder minimum candidate. These data span a 50+ year baseline. The majority of the data come from the Mount Wilson program, which obtained 107 observations of this star using the ``HKP-1" photometer from 1966–1977, and continued to obtain 1547 observations with the upgraded ``HKP-2" photometer from 1977–1995. The California Planet Search later picked up this target, obtaining 9 observations in June–September of 1997 with Keck-HIRES (``HIRES-1"), and again after the 2004 detector upgrade (``HIRES-2"), obtaining 103 spectra between 2004 and March 2020. \SHK{} values were measured following \citet{Isaacson2010}. There is no guarantee that the \SHK{} values across all 4 instruments (MW HKP-1, MW HKP-2, HIRES-1, HIRES-2) have absolute agreement. \citetalias{Baum2022} investigated the need for offsets between the various instruments and found that the Mount Wilson data (HKP-1 and HKP-2) appeared to agree without needing any offset. For the Keck-HIRES data, the \SHK{} values from HIRES-1 in 1997 \emph{do} appear to be higher than the later HIRES-2 \SHK{} values. However, the HIRES-1 data fail to establish a long enough time baseline for any conclusive evidence of an offset.  Further inspection by eye hints that the higher HIRES-1 \SHK{} values could be consistent with both the Mount Wilson and HIRES-2 values, occurring during a transition from higher, cycling activity in Mount Wilson to lower, flat activity in Keck-HIRES. Thus, no offsets were applied to the reported \SHK{} values in any of the 4 instruments in \citetalias{Baum2022}. This time series can be seen in the top panel of \autoref{fig:combined_data}.

\subsection[]{Additional Mount Wilson \SHK{} activity \citep{Olah2016}}
Upon investigating this star further, we were made aware of additional observations from the Mount Wilson program that took place between 1995 and 2002. The full Mount Wilson data (from 1966–2002) were not published, except for a figure in \citet{Olah2016}. We used a web application to scrape the data from Figure 2 of \citet{Olah2016} for HD 166620. As a result of the scraping process, the scraped data is reliable, but not exact. For example, due to the marker size, dense sampling, and long time baseline of the plot, the scraping procedure was forced to find the mean x and y positions over several points when markers overlapped. We compared the scraped data set with the \citetalias{Baum2022} data set during the overlapping time (1966–1995) and in general found excellent agreement in both time and \SHK{}. Given that \citetalias{Baum2022} has the exact data for their Mount Wilson observations, we opt to use the \citetalias{Baum2022} data for observations prior to June 4, 1995 (the final MW observation in that data set). We use the scraped MW data in \citet{Olah2016} for observations after that date. The new MW data therefore span from June 1995 to June 2002 and contains 119 (scraped) observations. The additional MW data from \citet{Olah2016} can be seen in the second panel of \autoref{fig:combined_data}. All activity data are presented in \autoref{tbl:activity}

\begin{deluxetable}{cccc}
\tablecaption{\label{tbl:activity} All \SHK{} activity data}
\tablehead{\colhead{BJD} & \colhead{Year} & \colhead{\SHK} & \colhead{Inst}}
\startdata
\input{activity_table.tex}
\enddata
\tablecomments{\autoref{tbl:activity} is published in its entirety in the machine-readable format.
      A portion is shown here for guidance regarding its form and content.}
\end{deluxetable}

\subsection{Photometry from the T4 Automated Photoelectric Telescope}
We acquired 1278 photometric observations of HD~166620 covering 17 observing 
seasons from 1993 to 2020 (we did not observe the star during the 2005 
through 2015 observing seasons).  The observations were all obtained with the 
T4 0.75~m automatic photoelectric telescope (APT) at Fairborn Observatory in 
southern Arizona.  The T4 APT is equipped with a single channel photometer 
that uses an EMI 9124QB bi-alkali photomultiplier tube to measure stellar 
brightness successively in the Str\"omgren $b$ and $y$ passbands.  

The observations of HD~166620 (star~d) were made differentially with respect 
to three nearby comparison stars (a, b, c).  Intercomparison of the six
combinations of differential magnitudes (d-a, d-b, d-c, c-a, c-b, and b-a)
reveals that only comparison star b (HD~166640) appears to be constant to
the limit of our precision.  Therefore, we present our results as differential
magnitudes in the sense star~d minus star~b, which we designate as d-b. 

To improve the photometric precision of the individual nightly observations, 
we combine the differential $b$ and $y$ magnitudes into a single $(b+y)/2$ 
``passband.''  The precision of a single observation with T4, as 
measured from pairs of constant comparison stars, typically ranges between 
0.0015~mag and 0.0020~mag on good photometric nights.  The T4 APT is described 
in \citet{Henry1999}, where further details of the telescope, precision photometer, 
and observing and data reduction procedures can be found.

We compute the standard deviations of the nightly observations for each observing season, which range from 0.00091 to 0.00197~mag, indicating little or no 
short-term variability within each observing season. We also compute the seasonal means and perform a frequency analysis of each individual observing season using the method of \citet{Vanicek1971}, which confirms the lack of any periodic variability.  \citet{Henry2022} show extensive examples of this method of period analysis.

\begin{figure*}
    \centering
    \includegraphics[width=\textwidth]{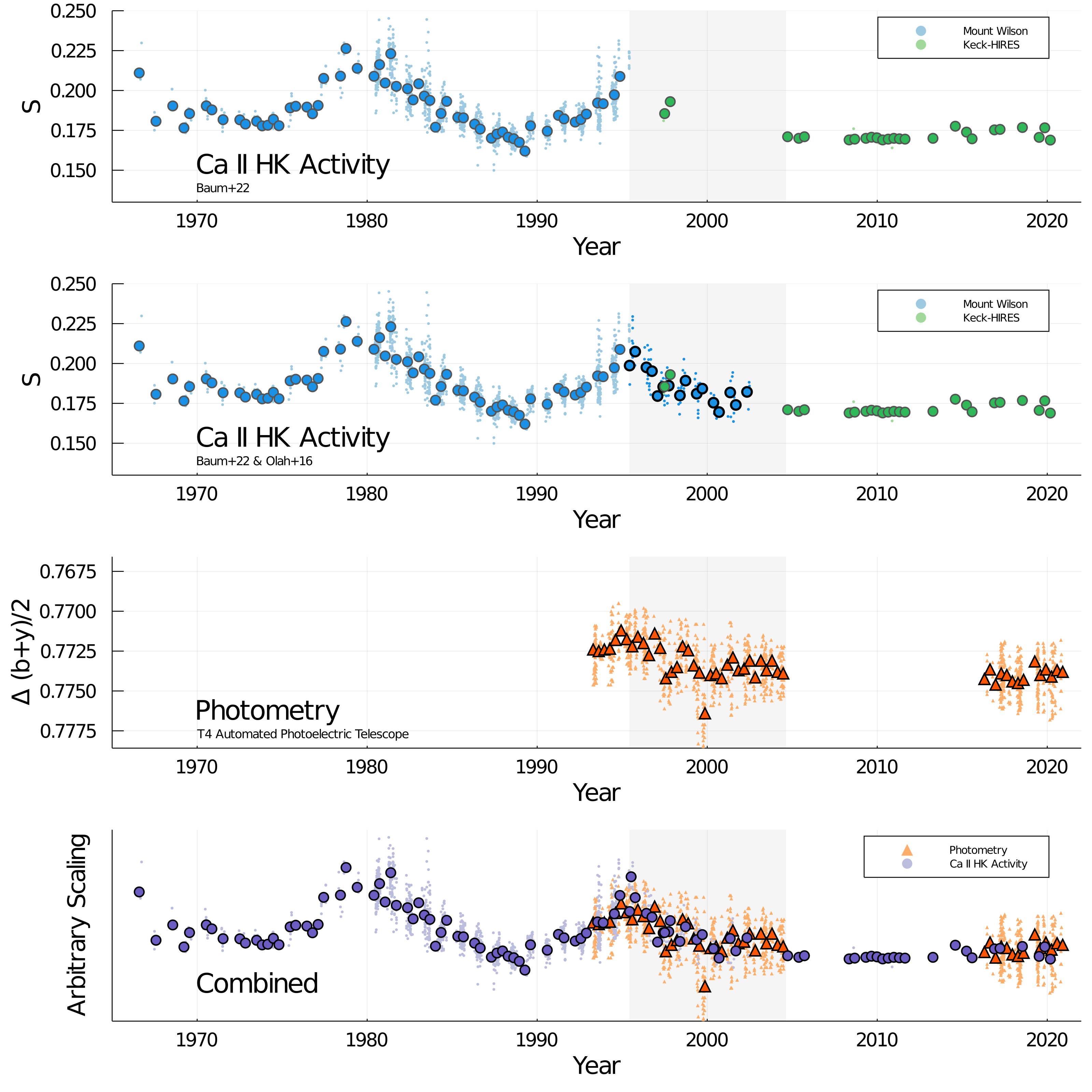}
    \caption{Photometry and \SHK{} time series for HD 166620 showing transition into Maunder Minimum. The top panel shows the \SHK{} time series from \citetalias{Baum2022} used to identify HD 166620 as a Maunder Minimum candidate. Blue points are data from the Mount Wilson program, green points are from Keck-HIRES. Large circles show the same data in 120-day bins. The gray shaded region indicates the gap between the Mount Wilson data and the post-2004 HIRES data (after the upgrade). The second panel shows the same data, but now includes the additional Mount Wilson data from \citet{Olah2016}, which fills in the previous gap and shows the clear transition from activity maximum to flat activity seen in Keck-HIRES. The third panel shows the T4 APT photometry, which similarly shows a transition from brighter to fainter during the transition period. Recent data (2016–) are at roughly the same magnitude as at the end of the transition. The bottom panel places the activity and photometry on the same scale by subtracting off the median of pre-2004 data, and multiplying the photometry by a scale factor of 10. The activity and photometry time series show remarkable agreement and unambiguously show a star transitioning from cycling to flat behavior.}
    \label{fig:combined_data}
\end{figure*}

The APT photometry can be seen in the third panel of \autoref{fig:combined_data} and are given in \autoref{tbl:photometry}.

\begin{deluxetable}{cccc}
\tablecaption{\label{tbl:photometry} T4 APT photometry}
\tablehead{\colhead{BJD} & \colhead{Year} & \colhead{$d-b$} & \colhead{Inst}}
\startdata
\input{photometry_table.tex}
\enddata
\tablecomments{\autoref{tbl:photometry} is published in its entirety in the machine-readable format.
      A portion is shown here for guidance regarding its form and content.}
\end{deluxetable}

\section{Analysis}
\subsection{Activity Time Series}
Considering only the \SHK{} time series from the combined \citetalias{Baum2022} and \citet{Olah2016} data, we see excellent agreement between all data sets and instruments. These data paint a picture of a star fully transitioning from cycling behavior to flat behavior, indicative of the beginning of a Maunder Minimum-like phase.

As a result, we can conclude that the HIRES-1 data (the pre-upgrade Keck-HIRES data in 1997) are consistently calibrated with both the HIRES-2 observations and all Mount Wilson observations, thus dispelling any doubts about potential offset errors between the Mount Wilson and Keck-HIRES data while also confirming that these are observations of the same star. 

We can then examine the activity level during its grand magnetic minimum as compared to the activity minima during cycling behavior. We identify two cycle minima prior to the transition period: the minimum occurring from 1971–1975 and the minimum occurring from 1986–1990. The mean activity level during the first minimum (1971.0–1975.0) is \SHK{}$ = 0.1810 \pm 0.0035$. The second minimum (1987.0–1991.0) has a mean activity level of \SHK{}$ = 0.1724 \pm 0.0058$. The period of flat activity (2004–) has a mean activity level of \SHK{}$=0.1705 \pm 0.0020$. Thus, the period of flat activity associated with the grand magnetic minimum is not significantly lower than what the star exhibits during regular cycling behavior. However, these values also suggest that the cycle minima decayed in the final cycles before the transition into grand magnetic minimum (and this is also suggested by the decaying cycle maxima in 1980 and 1995). Thus, it may be the case that we have caught this star at the end of a much larger interval of decaying cycles \footnote{This interpretation lends additional evidence that HD 4915 \citep{Shah2018} may soon be entering its own grand magnetic minimum.}. Without a longer time baseline prior to 1966, it is impossible to tell whether that is part of a larger trend. Either way, these results imply that the grand magnetic minimum is not the result of a qualitatively different surface magnetic field than is present during regular activity cycles.  We do note, however, that the variability in the grand minimum \emph{is} significantly lower than during either preceding cycle minimum.  In the Sun, \SHK~ in cycle minima are dominated by the roughly spatially uniform network, with occasional small active regions providing limited variability \citep[e.g.,][]{Milbourne2019}. This suggests in its grand minimum, HD 166620 may be even more dominated by network than the Sun in its cycle minima. 

We need to be careful though, as the \SHK~data from the Mount Wilson and Keck programs have different average noise levels.  \citet{Baliunas1995} gives $\sigma_S \approx 1.2\%$ from observations of their least variable stars. \citet{Wright04} give a similar value for their differential \SHK, but on the timescale of a single season, it seems this is an overestimate: the calculated $\sigma_S$ is less than this in 6 of the 9 seasons with more than one \SHK(Keck) measurement.  
The lowest seasonal $\sigma_S$ values approach 0.3\%; we adopt this as an upper limit to the seasonal noise for Keck.  If we subtract these estimates of the random noise on \SHK~in quadrature from the observed values, we have $\sigma_S$(min. 1) = 0.0027, $\sigma_S$(min. 2) = 0.0054, $\sigma_S$(grand) = 0.0019.  Even corrected for intrinsic noise differences, the variability in the grand minimum is still significantly lower than preceding minima; for example, an F-test gives 8.78$\times 10^{-6}$ chance that $\sigma_S$(grand) is drawn from the same distribution as $\sigma_S$(min. 1). 

\subsection[]{Combined Photometry \& \SHK}
We combined the APT photometry and \SHK{} onto a common scale in the bottom panel of \autoref{fig:combined_data}. To do so, we subtracted off the median of each data set prior to 2004, chosen to center the data on the cycling portion of the time series. We then multiplied the APT photometry by a scale factor of 10 to cover a similar range to \SHK, chosen arbitrarily by eye (scale factors of 8–12 all looked similarly consistent). The resulting combined time series shows excellent agreement between activity and brightness, showing that when the star transitioned from cycling to flat behavior, the star also became slightly fainter. Its mean brightness varied for the first several years over a range of 2 to 3 mmag. After 2004, the scatter in the mean magnitudes is only 0.35 mmag. Combined with the scarcity of strong rotationally modulated signals in either the activity or photometry, we see this as evidence that the star is not dominated during active periods by spots, but rather by faculae and a network of bright magnetic regions \citep[as seen on the sun, e.g.,][]{Milbourne2019} that lead to the star dimming as it decreases in activity. Similar correlations and interpretations have been made for other stars, e.g., \citet{Lockwood07}.

\subsection[]{Period Analysis of the HK Data}
We computed a floating-mean Lomb-Scargle periodogram for each season of data containing more than ten observations.  We searched for significant (FAP $\leq 5 \times 10^{-5}$) periodogram peaks near (within $\pm \approx$20\%) of the previously reported mean period \citep[$P_{\rm rot}$ = 42.4 days;][]{Donahue1996}.  Since double active longitudes are common, especially in less active stars \citep{Basri2018}, we also searched for significant periods within $\pm20$\% of $P_{\rm rot}/2$ = 21.2 days.  We compared results with computed harmonics and the data window function to discard aliases.  We found a total of nine seasons with significant periods, yielding an unweighted average (including doubled $P_{\rm rot}/2$ detections) of $\langle P_{\rm rot}\rangle = 45.06 \pm 4.07$ days.  The full range of detections spanned from 37.96 days to 50.99 days.  The scatter of $P_{\rm rot}$ values likely reflects a complex combination of surface differential rotation, plus activity growth and decay.  This updated $P_{\rm rot}$ and range should improve on the value given in \citet{Donahue1996}, as it includes more seasons (9 vs.7), while discarding some less certain previous $P_{\rm rot}$ values by using a more stringent FAP threshold.

\section{Discussion and Conclusions}

The dramatic nature and suspicious timing of the change in behavior of HD 166620 led \citetalias{Baum2022} to explore the possibility of a misidentification of stars between the two sets; and the presence of offsets among the HIRES and Mount Wilson data for a few stars raised questions about the strength of conclusions one can draw about the relative activity levels before and after this change.

The new Mount Wilson data here are contemporaneous with the pre-2004 HIRES data in \citetalias{Baum2022} and consistent with it, demonstrating that the mutual calibration there is robust, and that the HIRES data do show the end of the star's last activity cycle.

Our new photometry spans the pre- and post-2004 HIRES data, and shows qualitatively and quantitatively the same behavior, that is, a consistent and positive correlation between the star's optical brightness and Ca {\sc ii} H \& K activity level. A possible exception is in seasonal variability in \SHK, which is notably lower in the grand minimum.  Together, these photometric data span the three data sets of \citetalias{Baum2022}, removing any concern that the sets might not be of the same star or that large calibration offsets exist among them.  

We conclude that HD 166620 is the first unambiguous Maunder-Minimum analog, identified by its activity time series as it switched from a cycling to a flat-activity state.  Its activity history shows that, at least in this case, the average activity level in grand activity minima like the Maunder Minimum is not significantly lower than a star experiences when cycling, complicating efforts to identify such stars via their mean or instantaneous activity levels.  Lower variability in \SHK, though, may prove useful in diagnosing grand minima; more observed minima are needed to test this, however.

This also implies that the grand minimum is not the result of a qualitatively different surface magnetic field, for instance due to the collapse of the dynamo, but is essentially an ordinary minimum extended in time, perhaps with fewer residual active regions to explain the further reduced variability.  This is consistent with the fact that the Sun had a small number of sunspots during the Maunder Minimum \citep[e.g.,][]{Ribes1993} and that cosmogenic Be$^{10}$ continued to vary \citep[modulated by changing solar field topology, e.g.,][]{Beer1998} showing that the surface magnetic fields were still present then. 

We also note that the mean activity level of HD 166620 is $\logRHK=-5.03$ \citep{Brewer16}\footnote{$\logRHK=-5.07$ after correcting for metallicity following \citet{Saar2012}}, which is at about the level identified by \citet{Brandenburg2017} and \citet{Metcalfe2017} as the threshold for experiencing Maunder minimum behavior. Furthermore, taking $P_{\rm rot} = 42.4$ d \citep{Donahue1996} or our revised  $P_{\rm rot}$ = 45.06 days, the star's Rossby number is slightly larger than the Sun's \citep[2.08 or 2.21 vs. 1.99 using a turnover time $\tau_C$ from][]{Noyes1984}.  This is consistent with it being slightly older\footnote{Note that simple \citet{Barnes07}-style gyrochronology should still be valid for HD 166620, since with $T_{\rm eff} \approx 4970$K \citep{Brewer2016}, the star is warmer than the zone of spin-down ``stalling" \citep[see][]{Curtis2020}.} and less active than the Sun, and also consistent with their model.  
 
 The star was observed with ROSAT HRI in late 1996, just past the last cycle maximum; it displayed an X-ray luminosity of $\log L_X = 26.96$ \citep{Schmitt2004}. Adopting   $R=0.80 R_\odot$ \citep{Brewer2016}, this $L_X$ implies a surface flux of $F_X = 4.7\times10^4$ ergs cm$^{-2}$ s$^{-1}$, a value above the minimum level seen in dwarfs \citep[$F_X = 10^4$ ergs cm$^{-2}$s$^{-1}$;][]{Schmitt2004}, but less than the average solar level of $F_X \approx 1.3 \times 10^5$ ergs cm$^{-2}$ s$^{-1}$ \citep[converted from the average $L_X$ from][]{Judge2003}. This again, together with our updated rotation period of $P_{\rm rot} = 45.06$ days, is consistent with a star older and less active than the Sun, perhaps with a similarly faltering dynamo \citep{Metcalfe2017}.
 
\vspace{0.5cm}

We thank Travis Metcalfe for pointing us to the Ol\'ah et al.\ paper containing additional Mount Wilson for HD 166620, and for helpful comments.  We thank Howard Isaacson for discussions of Keck-HIRES activity measurements.

G.W. H acknowledges long-term support from Tennessee State University and the State of Tennessee through its Centers of Excellence Program. S.H.S. gratefully acknowledges support of HST grant HST-GO-16421.001-A and NASA XRP grant 80NSSC21K0607.

The Mount Wilson Observatory HK Project was supported by both public and private funds through the Carnegie Observatories, the Mount Wilson Institute, and the Harvard-Smithsonian Center for Astrophysics starting in 1966 and continuing for over 36 years. These data are the result of the dedicated work of O. Wilson, A. Vaughan, G. Preston, D. Duncan, S. Baliunas, and many others.

Some of the data presented herein were obtained at the W. M. Keck Observatory, which is operated as a scientific partnership among the California Institute of Technology, the University of California and the National Aeronautics and Space Administration. The Observatory was made possible by the generous financial support of the W. M. Keck Foundation.
The authors wish to recognize and acknowledge the very significant cultural role and reverence that the summit of Maunakea has always had within the indigenous Hawaiian community.  We are most fortunate to have the opportunity to conduct observations from this mountain.

This research has made use of NASA's Astrophysics Data System Bibliographic Services.

The Center for Exoplanets and Habitable Worlds 
and the Penn State Extraterrestrial Intelligence Center are
supported by the Penn State and the Eberly College of Science.

\bibliography{library,references}

\end{document}

%% file: activity_table.tex
2439342.80 & 1966.5926  & 0.211      & MW       \\
2439369.81 & 1966.6666  & 0.207      & MW       \\
2439392.80 & 1966.7296  & 0.2298     & MW       \\
2439669.80 & 1967.4885  & 0.1751     & MW       \\
2439670.79 & 1967.4912  & 0.1863     & MW       \\
\vdots & \vdots & \vdots & \vdots

%% file: photometry_table.tex
2449094.91 & 1993.29153 & 0.7746 & APT  \\
2449102.92 & 1993.31348 & 0.7724 & APT  \\
2449103.89 & 1993.31613 & 0.7739 & APT  \\
2449105.88 & 1993.3216  & 0.7728 & APT  \\
2449108.88 & 1993.32981 & 0.7746 & APT  \\
\vdots & \vdots & \vdots & \vdots

%% file: main.bbl
\begin{thebibliography}{}
\expandafter\ifx\csname natexlab\endcsname\relax\def\natexlab#1{#1}\fi
\providecommand{\url}[1]{\href{#1}{#1}}
\providecommand{\dodoi}[1]{doi:~\href{http://doi.org/#1}{\nolinkurl{#1}}}
\providecommand{\doeprint}[1]{\href{http://ascl.net/#1}{\nolinkurl{http://ascl.net/#1}}}
\providecommand{\doarXiv}[1]{\href{https://arxiv.org/abs/#1}{\nolinkurl{https://arxiv.org/abs/#1}}}

\bibitem[{{Baliunas} \& {Jastrow}(1990)}]{Baliunas90}
{Baliunas}, S., \& {Jastrow}, R. 1990, \nat, 348, 520, \dodoi{10.1038/348520a0}

\bibitem[{{Baliunas} {et~al.}(1995{\natexlab{a}}){Baliunas}, {Donahue}, {Soon},
  {Horne}, {Frazer}, {Woodard-Eklund}, {Bradford}, {Rao}, {Wilson}, {Zhang},
  {Bennett}, {Briggs}, {Carroll}, {Duncan}, {Figueroa}, {Lanning}, {Misch},
  {Mueller}, {Noyes}, {Poppe}, {Porter}, {Robinson}, {Russell}, {Shelton},
  {Soyumer}, {Vaughan}, \& {Whitney}}]{Baliunas95b}
{Baliunas}, S.~L., {Donahue}, R.~A., {Soon}, W.~H., {et~al.}
  1995{\natexlab{a}}, \apj, 438, 269, \dodoi{10.1086/175072}

\bibitem[{{Baliunas} {et~al.}(1995{\natexlab{b}}){Baliunas}, {Donahue}, {Soon},
  {Horne}, {Frazer}, {Woodard-Eklund}, {Bradford}, {Rao}, {Wilson}, {Zhang},
  {Bennett}, {Briggs}, {Carroll}, {Duncan}, {Figueroa}, {Lanning}, {Misch},
  {Mueller}, {Noyes}, {Poppe}, {Porter}, {Robinson}, {Russell}, {Shelton},
  {Soyumer}, {Vaughan}, \& {Whitney}}]{Baliunas1995}
---. 1995{\natexlab{b}}, \apj, 438, 269, \dodoi{10.1086/175072}

\bibitem[{{Barnes}(2007)}]{Barnes07}
{Barnes}, S.~A. 2007, ArXiv e-prints, 704

\bibitem[{{Basri} \& {Nguyen}(2018)}]{Basri2018}
{Basri}, G., \& {Nguyen}, H.~T. 2018, \apj, 863, 190,
  \dodoi{10.3847/1538-4357/aad3b6}

\bibitem[{{Baum} {et~al.}(2022){Baum}, {Wright}, {Luhn}, \&
  {Isaacson}}]{Baum2022}
{Baum}, A.~C., {Wright}, J.~T., {Luhn}, J.~K., \& {Isaacson}, H. 2022, \aj,
  163, 183, \dodoi{10.3847/1538-3881/ac5683}

\bibitem[{{Beer} {et~al.}(1998){Beer}, {Tobias}, \& {Weiss}}]{Beer1998}
{Beer}, J., {Tobias}, S., \& {Weiss}, N. 1998, \solphys, 181, 237,
  \dodoi{10.1023/A:1005026001784}

\bibitem[{{Brandenburg} {et~al.}(2017){Brandenburg}, {Mathur}, \&
  {Metcalfe}}]{Brandenburg2017}
{Brandenburg}, A., {Mathur}, S., \& {Metcalfe}, T.~S. 2017, \apj, 845, 79,
  \dodoi{10.3847/1538-4357/aa7cfa}

\bibitem[{{Brewer} {et~al.}(2016{\natexlab{a}}){Brewer}, {Fischer}, {Valenti},
  \& {Piskunov}}]{Brewer16}
{Brewer}, J.~M., {Fischer}, D.~A., {Valenti}, J.~A., \& {Piskunov}, N.
  2016{\natexlab{a}}, \apjs, 225, 32, \dodoi{10.3847/0067-0049/225/2/32}

\bibitem[{{Brewer} {et~al.}(2016{\natexlab{b}}){Brewer}, {Fischer}, {Valenti},
  \& {Piskunov}}]{Brewer2016}
---. 2016{\natexlab{b}}, \apjs, 225, 32, \dodoi{10.3847/0067-0049/225/2/32}

\bibitem[{{Curtis} {et~al.}(2020){Curtis}, {Ag{\"u}eros}, {Matt}, {Covey},
  {Douglas}, {Angus}, {Saar}, {Cody}, {Vanderburg}, {Law}, {Kraus}, {Latham},
  {Baranec}, {Riddle}, {Ziegler}, {Lund}, {Torres}, {Meibom}, {Aguirre}, \&
  {Wright}}]{Curtis2020}
{Curtis}, J.~L., {Ag{\"u}eros}, M.~A., {Matt}, S.~P., {et~al.} 2020, \apj, 904,
  140, \dodoi{10.3847/1538-4357/abbf58}

\bibitem[{{Donahue} {et~al.}(1995){Donahue}, {Baliunas}, {Soon}, \&
  {McMillan}}]{Donahue1995}
{Donahue}, R.~A., {Baliunas}, S.~L., {Soon}, W.~H., \& {McMillan}, F.~M. 1995,
  IAU Symposium, 176, 72

\bibitem[{{Donahue} {et~al.}(1996){Donahue}, {Saar}, \&
  {Baliunas}}]{Donahue1996}
{Donahue}, R.~A., {Saar}, S.~H., \& {Baliunas}, S.~L. 1996, \apj, 466, 384,
  \dodoi{10.1086/177517}

\bibitem[{{Eddy}(1976)}]{Eddy76}
{Eddy}, J.~A. 1976, Science, 192, 1189

\bibitem[{{Henry}(1999)}]{Henry1999}
{Henry}, G.~W. 1999, \pasp, 111, 845, \dodoi{10.1086/316388}

\bibitem[{{Henry} {et~al.}(2022){Henry}, {Fekel}, \& {Williamson}}]{Henry2022}
{Henry}, G.~W., {Fekel}, F.~C., \& {Williamson}, M.~H. 2022, \aj, 163, 180,
  \dodoi{10.3847/1538-3881/ac540b}

\bibitem[{{Henry} {et~al.}(1996){Henry}, {Soderblom}, {Donahue}, \&
  {Baliunas}}]{Henry96}
{Henry}, T.~J., {Soderblom}, D.~R., {Donahue}, R.~A., \& {Baliunas}, S.~L.
  1996, \aj, 111, 439

\bibitem[{{Isaacson} \& {Fischer}(2010)}]{Isaacson2010}
{Isaacson}, H., \& {Fischer}, D. 2010, \apj, 725, 875,
  \dodoi{10.1088/0004-637X/725/1/875}

\bibitem[{{Judge} {et~al.}(2003){Judge}, {Solomon}, \& {Ayres}}]{Judge2003}
{Judge}, P.~G., {Solomon}, S.~C., \& {Ayres}, T.~R. 2003, \apj, 593, 534,
  \dodoi{10.1086/376405}

\bibitem[{{Lockwood} {et~al.}(2007){Lockwood}, {Skiff}, {Henry}, {Henry},
  {Radick}, {Baliunas}, {Donahue}, \& {Soon}}]{Lockwood07}
{Lockwood}, G.~W., {Skiff}, B.~A., {Henry}, G.~W., {et~al.} 2007, \apjs, 171,
  260, \dodoi{10.1086/516752}

\bibitem[{{Metcalfe} \& {van Saders}(2017)}]{Metcalfe2017}
{Metcalfe}, T.~S., \& {van Saders}, J. 2017, \solphys, 292, 126,
  \dodoi{10.1007/s11207-017-1157-5}

\bibitem[{{Milbourne} {et~al.}(2019){Milbourne}, {Haywood}, {Phillips}, {Saar},
  {Cegla}, {Cameron}, {Costes}, {Dumusque}, {Langellier}, {Latham},
  {Maldonado}, {Malavolta}, {Mortier}, {Palumbo}, {Thompson}, {Watson},
  {Bouchy}, {Buchschacher}, {Cecconi}, {Charbonneau}, {Cosentino}, {Ghedina},
  {Glenday}, {Gonzalez}, {Li}, {Lodi}, {L{\'o}pez-Morales}, {Lovis}, {Mayor},
  {Micela}, {Molinari}, {Pepe}, {Piotto}, {Rice}, {Sasselov}, {S{\'e}gransan},
  {Sozzetti}, {Szentgyorgyi}, {Udry}, \& {Walsworth}}]{Milbourne2019}
{Milbourne}, T.~W., {Haywood}, R.~D., {Phillips}, D.~F., {et~al.} 2019, \apj,
  874, 107, \dodoi{10.3847/1538-4357/ab064a}

\bibitem[{{Mittag} {et~al.}(2019){Mittag}, {Schmitt}, {Metcalfe}, {Hempelmann},
  \& {Schr{\"o}der}}]{Mittag2019}
{Mittag}, M., {Schmitt}, J.~H.~M.~M., {Metcalfe}, T.~S., {Hempelmann}, A., \&
  {Schr{\"o}der}, K.~P. 2019, \aap, 628, A107,
  \dodoi{10.1051/0004-6361/201935654}

\bibitem[{{Noyes} {et~al.}(1984){Noyes}, {Hartmann}, {Baliunas}, {Duncan}, \&
  {Vaughan}}]{Noyes1984}
{Noyes}, R.~W., {Hartmann}, L.~W., {Baliunas}, S.~L., {Duncan}, D.~K., \&
  {Vaughan}, A.~H. 1984, \apj, 279, 763, \dodoi{10.1086/161945}

\bibitem[{{Ol{\'a}h} {et~al.}(2016){Ol{\'a}h}, {K{\H{o}}v{\'a}ri}, {Petrovay},
  {Soon}, {Baliunas}, {Koll{\'a}th}, \& {Vida}}]{Olah2016}
{Ol{\'a}h}, K., {K{\H{o}}v{\'a}ri}, Z., {Petrovay}, K., {et~al.} 2016, \aap,
  590, A133, \dodoi{10.1051/0004-6361/201628479}

\bibitem[{{Ribes} \& {Nesme-Ribes}(1993)}]{Ribes1993}
{Ribes}, J.~C., \& {Nesme-Ribes}, E. 1993, \aap, 276, 549

\bibitem[{{Saar} \& {Baliunas}(1992)}]{Saar92}
{Saar}, S.~H., \& {Baliunas}, S.~L. 1992, in ASP Conf. Ser. 27: The Solar
  Cycle, ed. K.~L. {Harvey}, 150--167

\bibitem[{{Saar} \& {Testa}(2012)}]{Saar2012}
{Saar}, S.~H., \& {Testa}, P. 2012, in Comparative Magnetic Minima:
  Characterizing Quiet Times in the Sun and Stars, ed. C.~H. {Mandrini} \&
  D.~F. {Webb}, Vol. 286, 335--345, \dodoi{10.1017/S1743921312005066}

\bibitem[{{Schmitt} \& {Liefke}(2004)}]{Schmitt2004}
{Schmitt}, J.~H.~M.~M., \& {Liefke}, C. 2004, \aap, 417, 651,
  \dodoi{10.1051/0004-6361:20030495}

\bibitem[{{Shah} {et~al.}(2018{\natexlab{a}}){Shah}, {Wright}, {Isaacson},
  {Howard}, \& {Curtis}}]{Shah2018}
{Shah}, S.~P., {Wright}, J.~T., {Isaacson}, H., {Howard}, A., \& {Curtis},
  J.~L. 2018{\natexlab{a}}, ArXiv e-prints.
\newblock \doarXiv{1801.09650}

\bibitem[{{Shah} {et~al.}(2018{\natexlab{b}}){Shah}, {Wright}, {Isaacson},
  {Howard}, \& {Curtis}}]{Shah18}
{Shah}, S.~P., {Wright}, J.~T., {Isaacson}, H., {Howard}, A.~W., \& {Curtis},
  J.~L. 2018{\natexlab{b}}, \apjl, 863, L26, \dodoi{10.3847/2041-8213/aad40c}

\bibitem[{{Van{\'\i}{\v{c}}ek}(1971)}]{Vanicek1971}
{Van{\'\i}{\v{c}}ek}, P. 1971, \apss, 12, 10, \dodoi{10.1007/BF00656134}

\bibitem[{{Wright}(2004)}]{Wright04b}
{Wright}, J.~T. 2004, \aj, 128, 1273, \dodoi{10.1086/423221}

\bibitem[{{Wright} {et~al.}(2004){Wright}, {Marcy}, {Butler}, \&
  {Vogt}}]{Wright04}
{Wright}, J.~T., {Marcy}, G.~W., {Butler}, R.~P., \& {Vogt}, S.~S. 2004, \apjs,
  152, 261

\end{thebibliography}
